\documentclass[12pt,a4paper]{article}
\usepackage{graphicx,fancyhdr}
\usepackage{cite,citesort}
\usepackage{color}
\begin{document}

\pagenumbering{arabic}

\vspace{-40mm}

\begin{center}
\Large\textbf{Magnetotransport properties of iron microwires fabricated
by focused electron beam induced autocatalytic growth}\normalsize \vspace{8mm}

\large F. Porrati$^1$, R. Sachser$^1$, M.-M. Walz$^2$, F. Vollnhals$^2$, H.-P. Steinr\"{u}ck$^2$, H.~Marbach$^2$ and M. Huth$^1$

\vspace{5mm}

\small\textit{1. Physikalisches Institut, Goethe-Universit\"at,
Max-von-Laue-Str.~1, D-60438 Frankfurt am Main, Germany}

\small\textit{2. Lehrstuhl f\"{u}r Physikalische Chemie II,
Universit\"{a}t Erlangen-N\"{u}rnberg, Egerlandstrasse~3, D-91058 Erlangen, Germany}

\end{center}

\vspace{0cm}
\begin{center}
\large\textbf{Abstract}\normalsize
\end{center}

We have prepared iron microwires in a combination of focused electron beam induced deposition (FEBID) and autocatalytic growth from the iron pentacarbonyl, $Fe(CO)_5$, precursor gas under UHV conditions. The electrical transport properties of the microwires were investigated and it was found that the temperature dependence of the longitudinal resistivity ($\rho_{xx}$) shows a typical metallic behaviour with a room temperature value of about 88~$\mu\Omega~cm$. In order to investigate the magnetotransport properties we have measured the isothermal Hall-resistivities in the range between 4.2~K and 260~K. From these measurements positive values for the ordinary and the anomalous Hall coefficients were derived. The relation between anomalous Hall resistivity ($\rho_{AN}$) and longitudinal resistivity is quadratic, $\rho_{AN}$~$\sim$~$\rho_{xx}^2$, revealing an intrinsic origin of the anomalous Hall effect. Finally, at low temperature in the transversal geometry a negative magnetoresistance of about 0.2~$\%$ was measured.

\begin{center}
\large\textbf{1. Introduction}\normalsize
\end{center}

The ability to control matter down to the nanometer scale is basic for the development of new artificial materials and devices in nanotechnology. Focused electron beam induced deposition (FEBID) is an emerging direct writing technique used to fabricate samples down to the nanometer scale~\cite{utke,van dorp,furuya}. Within this technique the fabrication of the samples takes place by using the adsorbed molecules of a precursor gas injected in a scanning electron microscope (SEM): by interaction with the electron beam of the SEM the adsorbed molecules dissociate into a volatile component, eventually pumped away and into a non-volatile one, which constitutes the sample, also called deposit. By rastering with the electron beam over the area of interest structures with the desired shape can be written. The availability of a large number of precursors~\cite{utke,van dorp} enables the fabrication of structures with variable chemical composition. The spectrum of target materials ranges from insulators to semiconductors, to metals, to superconductors~\cite{utke}. This is combined with the excellent lateral resolution intrinsic to the FEBID process and thus allows for the fabrication of artificial and tunable solid state model systems~\cite{fab_2}. Magnetic materials mainly prepared from $Co_2(CO)_8$ and $Fe(CO)_5$, are of particular relevance in this regard~\cite{lau,kunz,shimojo}. Magnetic nanostructures have been fabricated and proposed as high resolution magnetic tips\cite{utke_2}, domain wall based memory devices\cite{fernandez_2} and magnetic sensors\cite{boero,gabureac}.

Despite the growing interest for magnetic materials fabricated by FEBID, only very few studies have been performed to investigate their electrical\cite{lau,lavrijsen} and magnetotransport properties\cite{fernandez,boero,gabureac}. This is due to the difficulty to obtain pure materials, which are often required for applications.
Recently, much effort has been devoted to improve the metal content of FEBID materials~\cite{botman}, which often consists of metal clusters embedded in a carbonaceous matrix~\cite{lau,takeguchi,utke_2}. One strategy to obtain higher metal content deposits is to optimize the deposition parameters~\cite{shimojo,lau,huth,lavrijsen,fernandez}. Furthermore, various purification techniques can be used, as recently reviewed by Botman et al.~\cite{botman}. These includes the deposition on hot substrates\cite{mulders}, post-growth annealing~\cite{botman_2} also combined with the dosage of a reactive gas~\cite{schirmer}, post-growth electron irradiation~\cite{fab} and the deposition in mixed gas atmosphere\cite{folch}. Such techniques have been applied for FEBID fabrication of iron-based nanocomposites prepared from $Fe(CO)_5$ (post deposition heating) and $Fe_2(CO)_9$ (additional dosage of water during deposition), and yielded metal contents as high as 70$\%$ and 75$\%$\cite{takeguchi,lavrijsen}, respectively. A further reduction of the contamination level in deposits from $Fe(CO)_5$ has been achieved by performing FEBID under ultra-high vacuum (UHV) conditions~\cite{lukas}. With this approach the level of carbon and oxygen contaminations is greatly reduced, leading to deposits with a metal content significantly higher than 90$\%$~\cite{lukas}. Working under UHV also allows for a novel two-step protocol to generate clean iron structures, namely focused electron beam induced surface activation (FEBISA)~\cite{walz}. In an exemplary study, a $\textit{SiO$_{x}$}$ sample is in a first processing step locally activated by the irradiation with a focused electron beam. In a second step the corresponding activated patterns are exposed to the precursor $Fe(CO)_5$, which is then catalytically decomposed at the electron irradiated positions resulting in pure iron deposits. These iron deposits continue to grow autocatalytically as long as the precursor is supplied. Remarkably, this process proceeds already at room temperature in UHV. The process is even more effective, if the first step is a true FEBID step, i.e., if the precursor gas is already present during irradiation with electrons. Herein, we report the electrical and the magnetotransport properties of iron microstructures grown using this latter approach.

\begin{center}
\large\textbf{2. Experimental details}\normalsize
\end{center}

We have fabricated iron microwires in a two-chamber UHV system (Omicron Multiscan Lab) with a base pressure below $2\times10^{-10}$~mbar. The chamber houses an UHV-compatible SEM with a resolution better than 3~nm at a beam current of 400~pA and acceleration voltage of 15~kV. Composition analysis can be performed in-situ by local Auger electron spectroscopy (AES) with a resolution better than 10~nm employing an hemispherical electron energy analyzer. As a precursor we used iron pentacarbonyl with a purity of 99.5~$\%$ from the Sigma-Aldrich company. The dosage of the precursor gas was performed through a dosing nozzle with an inner diameter of 3~mm, positioned approximately 12~mm away from the sample surface leading to an estimated local pressure at the surface of  $9\times10^{-6}$~mbar, which corresponds to an enhancement by a factor of 30 as compared to the background pressure.

The transport measurements were performed several days after the sample preparation, with the samples stored at ambient conditions for some days. The measurements were performed in a variable-temperature insert mounted in a $^4$He cryostat equipped with a 9 T superconducting solenoid. The temperature ranged between 1.8~K  and 265~K. A current of 10~$\mu$A, for a resulting current density of about 1.25$\times~10^8 A~m^{-2}$, was applied to the sample by using a Keithley Sourcemeter 2400. In order to measure the voltage during the four-probe resistance and the Hall measurements we employed an Agilent 34420A nanovoltmeter.

\begin{center}
\large\textbf{3. Results}\normalsize
\end{center}

\begin{center}
\large\textbf{3.1 Fabrication of the microwires}\normalsize
\end{center}

\begin{figure}\center{\includegraphics[width=10cm]{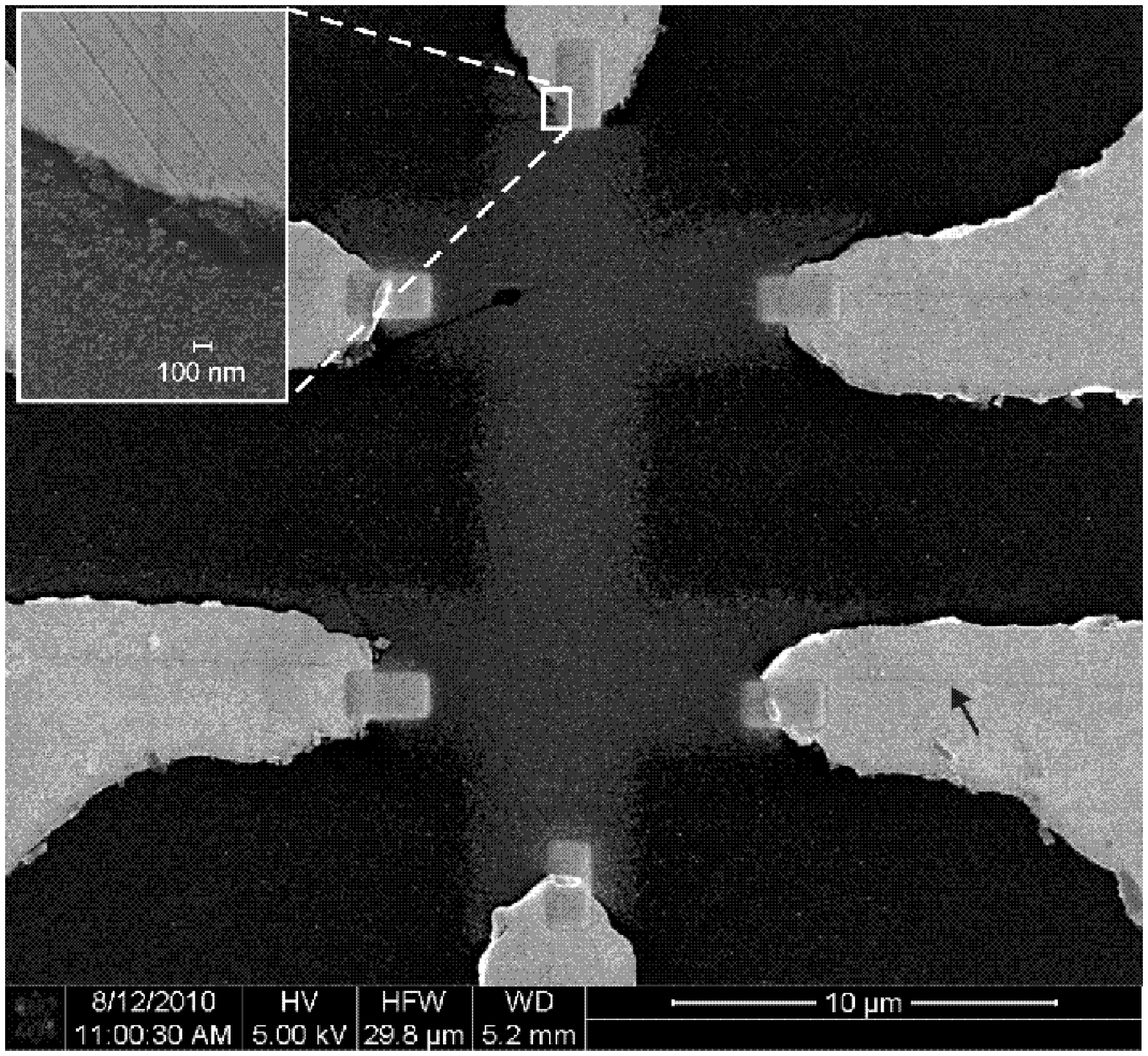}}
\caption{SEM picture of a microwire prepared for 4-probe transport measurements. The sample is constituted of Fe crystallites with a typical size of ca.~50~nm with a standard deviation of 11~nm (see inset). The thickness of the sample is ca.~30~nm, as measured by AFM. The contact electrodes were prepared by standard UV photolitography with 120~nm Au/Cr. Since a gap was present between microwires and electrodes (see inset), bridge contacts made of low-resistance W-C-Ga-based ion-beam-induced deposits using the precursor $W(CO)_6$ were added. The image shown in the inset was taken before writing the bridge contacts. The black arrow on the lower right electrode marks a narrow line originated in the FEBID process, see text for details.}
\label{microwire}\end{figure}

In Fig.~\ref{microwire} a (double-cross shaped) iron microwire structure generated by FEBID and successive autocatalytic growth is depicted. This iron microwire was prepared at room temperature on a Si(p-doped)/SiO$_2$(300~nm) substrate; the electron beam current was 400~pA at an energy of 15~keV and the accumulated line dose was 1.9~$\mu$C/cm. During electron irradiation, the background pressure of $Fe(CO)_5$ was $3\times10^{-7}$~mbar. After irradiation with electrons, the pressure was kept at the same level for additional 210 minutes to allow for continued autocatalytic growth at the deposited iron nuclei. The width of the iron lines is roughly 4~$\mu$m, which is expected from the simulated
exit area of backscattered electrons on $SiO_2$. The length of
the vertical and the horizontal irradiated lines was 45$\mu$m (i.e., significantly longer than required) in order to bridge the gaps between the gold contacts and to ensure a sufficient electrical contact in all cases. The original path of the electron beam can be recognized as narrow dark lines visible on the electrodes (indicated in Fig.~\ref{microwire} by an arrow at the lower right gold contact). These narrow lines originate
from the conventional FEBID process on the gold contact. A comparison with the much
wider microlines on the $SiO_2$ substrate shows that the autocatalytic growth is much less
efficient on the Au contact. This is either due to the much smaller backscattering area on
Au (due to the higher density of Au), a shorter residence time and thus shorter surface
diffusion length of the precursor molecules on Au or to a lower catalytic activity of the
deposited iron atoms due to their interaction with gold. All these effects would lead to
the observed negligible autocatalytic growth process on the Au contacts. The sensitivity to the nature of the local substrate is also evident when inspecting the gap between the Au contact and the granular Fe wire, as depicted in the zoomed out Fig.~\ref{microwire}. Local Auger measurements, as it can be compared in Ref.~\cite{schirmer}, reveal the presence of chromium in the region of the gap stemming from the buffer layer, effectively inhibiting the growth of a sufficiently thick Fe wire via autocatalytic effects in this region.

\begin{figure}\center{\includegraphics[width=14cm]{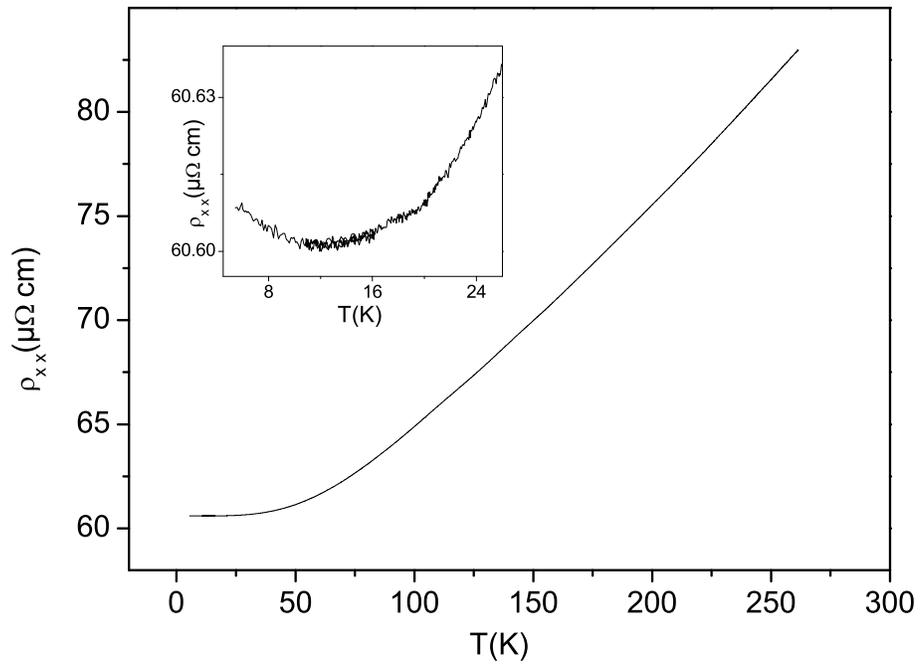}}
\caption{Temperature dependence of the longitudinal resistivity $\rho_{xx}$. The behavior is that of a metal. Below 12~K the resistivity slightly increases, see inset.}
\label{resistivity}\end{figure}

\begin{center}
\large\textbf{3.2 Magnetotransport measurements}\normalsize
\end{center}
In Fig.~\ref{resistivity} we plot the temperature dependence of the longitudinal resistivity for a typical microwire. The behavior is that of a metal with a resistivity of 84~$\mu\Omega~cm$ at 260~K, about a factor 8 higher than the value for pure bulk iron. An increase of the resistivity by reducing the thickness is expected from the surface scattering theory of Fuchs-Sondheimer~\cite{sondheimer}. At low temperature, see inset in Fig.~\ref{resistivity}, the resistivity shows a minimum; a similar effect was also observed in epitaxially grown Fe films~\cite{rubinstein,liu,sangiao} and was explained in terms of weak electron-localization and/or electron-interaction effects~\cite{rubinstein}.

\begin{figure}\center{\includegraphics[width=14cm]{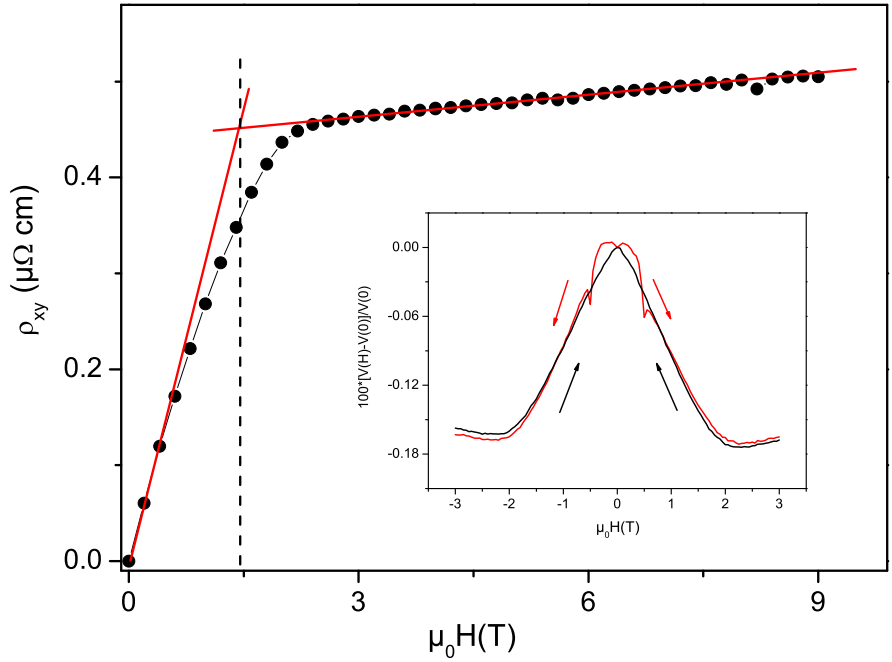}}
\caption{Hall resistivity $\rho_{xy}$ measured at 12~K for the microwire of Fig.\ref{microwire}. The slopes at high magnetic field determine the ordinary Hall coefficient. The intersection of the y axis with the extrapolation of the resistivity large fields towards zero field determines the anomalous Hall coefficient. The intersection of the slopes at low and high fields gives the saturation magnetization M$_s$ for known demagnetizing factor N (see text for details). Inset: Transverse magnetoresistance for magnetic field perpendicular to the surface of the microwire.}
\label{hall_effect}\end{figure}

In Fig.~\ref{hall_effect} we display the result of an Hall resistivity measurement performed at 12~K. The Hall resistivity, $\rho_{xy}$, is given by the sum of the ordinary and the anomalous Hall effects, $\rho_{xy}=\rho_{OR}+\rho_{AN}~$, with $\rho_{OR}=R_0B$ and $\rho_{AN}=R_s\mu_0M$, where $\textit{B}$ is the magnetic induction, $\textit{M}$ the magnetization of the wire,  $R_0$ and $R_s$ the ordinary and the anomalous Hall coefficients, respectively. The demagnetizing factor of the microwires used in our investigations is $N\approx1$~\cite{hubert}.
Therefore, the Hall resistivity becomes $\rho_{xy}=R_0\mu_0H+R_s\mu_0M$, $\textit{H}$ being the applied magnetic field, and the Hall coefficients can be extracted directly from the plot of Fig.~\ref{hall_effect}. For the ordinary Hall coefficient we find $R_0=7.0\cdot10^3 \mu\Omega\cdot cm\cdot T^{-1}$, which can be compared to the bulk value of polycrystalline iron films, $R_0=2.3\cdot10^3 \mu\Omega\cdot cm\cdot T^{-1}$\cite{raeburn}, and to that of epitaxially grown thin films, $R_0=5\cdot10^3 \mu\Omega\cdot cm\cdot T^{-1}$\cite{rubinstein}, measured at room temperature. From Fig.~\ref{hall_effect} we extract the value of the saturation magnetization to be $M_s$=1.47~T, which is smaller than 2.1~T of bulk Fe. The trend is similar to the one measured in sputtered polycrystalline Fe thin films where the saturation magnetization decreases with the thickness\cite{kim}.

In the inset of Fig.~\ref{hall_effect} we show the magnetoresistance measured for a magnetic field normal to the surface of the microwire. The magnetoresistance is negative and decreases with increasing magnetic field because the scattering probability decreases as more magnetic moments align to the magnetic field. For magnetic fields higher than the saturation field, the magnetoresistance varies as $H^2$. By increasing or decreasing the magnetic field from zero towards saturation, the magnetoresistance first slightly increases reaching a maximum at around 0.2~T and then it drops at about 0.5~T. This feature generates a small hysteresis, which shows the presence of two metastable states involved during the magnetization process.

\begin{figure}\center{\includegraphics[width=16cm]{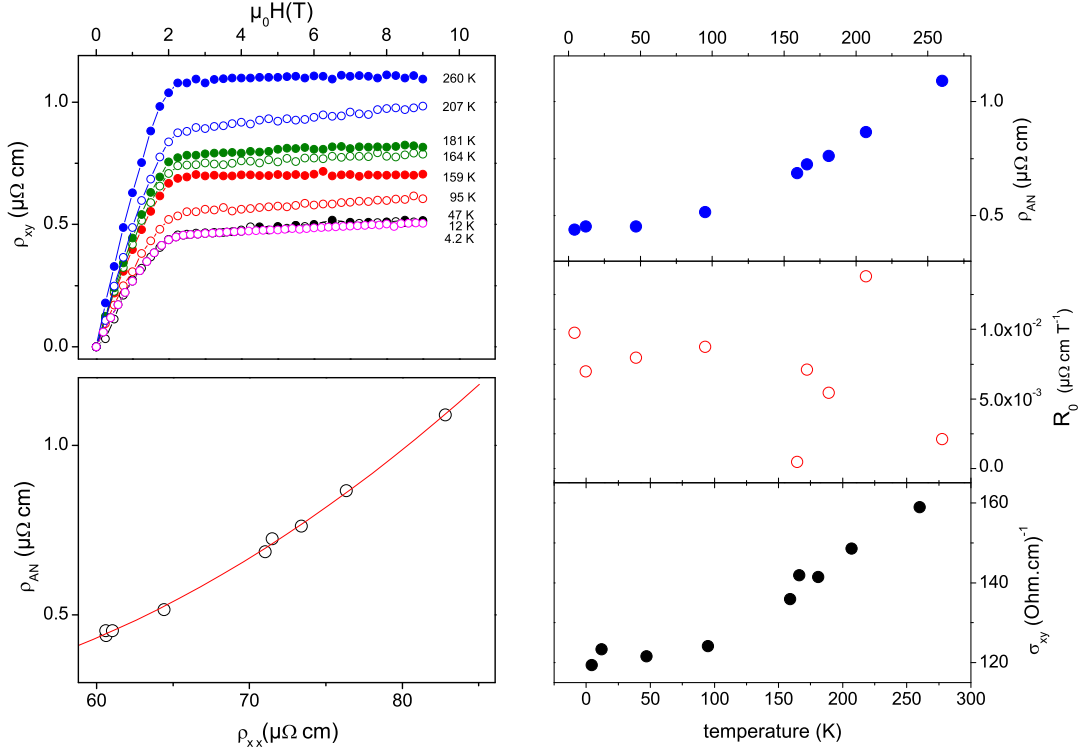}}
\caption{a) Hall resistivity vs.~applied magnetic field at various temperatures. b) Anomalous Hall resistivity, $\rho_{AN}$, vs. longitudinal resistivity, $\rho_{xx}$. The red line is the quadratic fit to the data: $\rho_{AN}$~$\sim$~$\rho_{xx}^2$. c) $\rho_{AN}$ vs. temperature. d) Ordinary Hall coefficient $R_0$ vs.~temperature. e) Anomalous Hall conductivity $\sigma_{xy}$  vs.~temperature. }
\label{anomalous}\end{figure}

In Fig.~\ref{anomalous}a) we report the results of the Hall resistivity measurements performed between 4.2~K and 260~K. From the analysis of the isothermal we find that, both, the ordinary and the anomalous Hall coefficients are positive in the temperature-range investigated (see figures~\ref{anomalous}d) and ~\ref{anomalous}c)). The ordinary Hall coefficient of our microwires is quite temperature independent, see Fig.~\ref{anomalous}d). Note that we attribute the deviations at 159~K, 206~K and 260~K to the difficulty to keep the temperature constant during the measurement which complicated the appropriate subtraction of spurious longitudinal contributions in the Hall voltage. The temperature independency of $R_0$ is also consistent with the data of Ref.~\cite{raeburn}. The anomalous Hall resistivity increases with the square of the longitudinal resistivity, $\rho_{AN}\propto\rho_{xx}^2$ (Fig~\ref{anomalous}b)), which excludes skew scattering as mechanism as a possible reason for the anomalous Hall effect. In Fig.~\ref{anomalous}e) we plot the temperature dependence of the anomalous Hall conductivity, defined as $\sigma_{xy}\simeq\rho_{xy}/\rho_{xx}{^2}$. The values of the conductivity ranges from 119~$\Omega^{-1} cm^{-1}$ to 158~$\Omega^{-1} cm^{-1}$; for a comparison with the literature see discussion below.

\begin{center}
\large\textbf{4. Discussion}\normalsize
\end{center}

One major problem of deposits prepared by FEBID is the formation of an unwanted additional deposit around the chosen deposition area. The lateral size of this co-deposit (or halo), whose existence has been reported for samples grown from a variety of precursors~\cite{lau,boero,utke,hochleitner}, can reach some micrometers depending on the material and on the electron beam parameters~\cite{boero}. The halo is due to backscattered electrons (BSE) and the secondary electrons (SE) generated from them~\cite{utke,van dorp,boero,hochleitner}. The presence of the halo is detrimental because it can affect the magnetic, the electrical and the magnetotransport measurements performed on the deposits\cite{boero,zhang,gopal}. The influence on the magnetic properties depends strongly on the precursor used. For example, it has been reported that nanowires grown from the $Co_2(CO)_8$ precursor gas have a quite extended co-deposit area~\cite{boero,fernandez_2}, with only the central part of the sample being ferromagnetic, indicating an inhomogeneous chemical composition~\cite{fernandez_2}. With the FEBID-based approach used here,
the efficient autocatalyic growth after initial electron irradiation leads to the
formation of an Fe layer with quite uniform thickness in the directly irradiated region plus in the region where a sufficient number of backscattered electrons exit the surface again. This is due to the fact that the majority of precursor molecules impinging on the surface decompose autocatalytically in the close vicinity of
their impact point. As a result, the autocatalytic growth of clean iron proceeds in the
whole area usually affected by the proximity effects. In this way the generation
of a chemically homogeneous iron microwires is realized.

The deposited iron microwires display metallic and ferromagnetic behavior. At room temperature the resistivity is about 88~$\mu\Omega~cm$, which is the lowest, to our knowledge, measured for iron deposits from $Fe(CO)_5$~\cite{shimojo,botman}. Furthermore, this resistivity is of the same order of magnitude of the one measured for cobalt nanowires~\cite{fernandez}, which is the lowest for FEBID deposits from carbonyl precursors in general. This is remarkable since the structures were exposed to ambient conditions for some days and are expected to be partly oxidized. For a comparison with samples prepared by IBID, which are often used because of their highly conduction behavior, note that tungsten deposits from $W(CO)_6$ have a typical resistivity of $\rho$ = 200~$\mu\Omega~cm$. Finally, it has to be mentioned that comparable resistivity values to the one obtained in the present work can be obtained by using carbon free precursors gases~\cite{botman,utke_4}.

In general, the results of the Hall resistivity measurements and the magnetoresistance measurements performed in our work, are in agreement with the investigations on iron thin films reported in the literature. In particular, we find that the anomalous resistivity has a quadratic dependence with the longitudinal resistivity. This dependence rules out skew-scattering as the mechanism at the origin of the AHE (linearity), but not the side jump effect (quadratic). It remains to understand if the effect is intrinsic (due to the band structure) or extrinsic (side jump). To answer this question we have to consider the anomalous Hall conductivity ($\sigma_{xy}\simeq\rho_{xy}/\rho_{xx}{^2}$). According to a recent unified theory valid for multiband ferromagnetic metals with dilute impurities, there exists a crossover between the intrinsic to the extrinsic anomalous Hall effect as a function of the longitudinal conductivity and the anomalous conductivity~\cite{onoda}. In the limit of highly conductive metals the anomalous conductivity is proportional to the longitudinal conductivity ($\sigma_{xy}\propto\sigma_{xx}$). In this region the AHE is extrinsic and its origin is the skew scattering effect. The intrinsic-to-extrinsic crossover takes place for lower conductivities ($\sigma_{xx}=10^4-10^6~\Omega^{-1}\cdot$~cm$^{-1}$), where the anomalous conductivity is constant with typical values of $\sigma_{xy}=10^2-10^3~\Omega^{-1}\cdot$~cm$^{-1}$\cite{miyasato,sangiao}. The values present in the literature for iron whiskers (1032~$\Omega^{-1} cm^{-1}$)~\cite{dherr} and 1~$\mu$m iron thin films (1000~$\Omega^{-1} cm^{-1}$)~\cite{miyasato} belong to this region (moderately dirty metals). Finally, in the dirty metal limit ($\sigma_{xx}<10^4~\Omega^{-1}\cdot$~cm$^{-1}$) $\sigma_{xy}\propto\sigma_{xx}^{1.6}$, which is found in the hopping transport regime\cite{miyasato}. The anomalous conductivity in our experiment is between 119~$\Omega^{-1}\cdot$~cm$^{-1}$ and 159~$~\Omega^{-1}\cdot$~cm$^{-1}$, see Fig.~\ref{anomalous}e). Correspondingly the longitudinal conductivity is in between 1.65$\cdot10^4~\Omega^{-1}\cdot$~cm$^{-1}$ and 1.21$\cdot10^4~\Omega^{-1}\cdot$~cm$^{-1}$. With respect to Refs.~\cite{onoda,sangiao} our samples have to be classified as being in the moderately dirty regime where the anomalous effect is intrinsic. It is interesting to note that the values of the conductivity measured in our experiment are in the same range as those reported recently for 2.0-2.5~nm thick Fe thin film epitaxially grown on MgO (001)~\cite{sangiao}.

\begin{center}
\large\textbf{5. Conclusions}\normalsize
\end{center}

We have prepared microwires using a novel fabrication technique, which mainly relies on the autocatalytic growth of clean iron structures from $Fe(CO)_5$. We have investigated the electrical- and magneto-transport properties of the microwires by means of Hall resistivity and magnetoresistance measurements. It turns out that the microwires are metallic with a resistivity of about 88~$\mu\Omega~cm$, which is among the lowest reported for samples prepared by FEBID from carbonyl precursors. The magnetotransport behavior of the microwires are comparable with those of iron thin films reported in the literature. In particular, we find that the magnetization saturation is $M_s$=1.47~T and the ordinary and anomalous Hall coefficients are positive in the temperature range investigated. The anomalous Hall resistivity scales quadratically with the longitudinal resistivity, which points to an intrinsic origin of the anomalous Hall effect. This conclusion is supported by the analysis of a recent unified theory for multiband ferromagnetic metals.

Finally, it should be mentioned that with the FEBID-based approach used here the
fabrication of pure iron line structures with line widths well below 100~nm is possible~\cite{walz}. Therefore it appears feasible to improve the linewidth of the iron structures to the nanoscale, which is required in spintronic and magneto recording technology, and which might make focused electron beam induced processing the technique of choice to prepare sub-100~nm periodic magnetic nanostructures used to study dipolar coupling effects~\cite{fab_4}.

\begin{center}
\large\textbf{Acknowledgments}\normalsize
\end{center}

The authors aknowledge financial support by the \textit{NanoNetzwerkHessen (NNH)}, the \textit{Bundesministerium f\"ur Bildung und Forschung (BMBF)} under grant no. 0312031C,
the Beilstein-Institut, Frankfurt/Main, Germany, within the research collaboration \emph{NanoBiC}, and the \textit{Deutsche Forschungsgemeinschaft} through grant MA 4246/1-2. In addition the support by COST action CM0601, COST action D41 and by the \textit{Cluster of Excellence Engineering of Advanced Materials (EAM)} is greatly acknowledged.

\newpage

\end{document}